\PassOptionsToPackage{dvipsnames}{xcolor}
\documentclass[10pt, a4paper, twocolumn]{article}

\usepackage[all]{nowidow}
\usepackage{xcolor}
\usepackage{subcaption}
\usepackage{hyperref}
\usepackage{acronym}
\usepackage{balance}
\usepackage[all]{nowidow}
\usepackage[capitalise,noabbrev]{cleveref}
\usepackage{authblk}

\usepackage{todonotes}
\newcommand{\free}{\vspace{1em}\noindent}

\date{}
\begin{document}

\title{Efficiently Detecting Performance Changes \\in FaaS Application Releases \thanks{This is the author copy of the paper published in the 9th International Workshop on Serverless Computing (WoSC '23) \textsuperscript{\textcopyright}2023 ACM (DOI: 10.1145/3631295.3631395). Aside from the formatting, it is identical with the official IEEE version.}}

\author{Martin~Grambow$^1$,
		Tim~Dockenfuß$^1$,
		Trever~Schirmer$^1$,
		Nils~Japke$^1$,
    and~David~Bermbach$^1$}
				
\affil{$^1$Mobile Cloud Computing Research Group, TU Berlin \\Berlin, Germany. \\E-mail: \{mg,td,ts,nj,db\}@mcc.tu-berlin.de}

\maketitle

\begin{abstract}
    The source code of Function as a Service (FaaS) applications is constantly being refined. 
    To detect if a source code change introduces a significant performance regression, the traditional benchmarking approach evaluates both the old and new function version separately using numerous artificial requests.
				
    In this paper, we describe a wrapper approach that enables the Randomized Multiple Interleaved Trials (RMIT) benchmark execution methodology in FaaS environments and use bootstrapping percentile intervals to derive more accurate confidence intervals of detected performance changes.
    We evaluate our approach using two public FaaS providers, an artificial performance issue, and several benchmark configuration parameters.
    We conclude that RMIT can shrink the width of confidence intervals in the results from $10.65\%$ using the traditional approach to $0.37\%$ using RMIT and thus enables a more fine-grained performance change detection.
\end{abstract}

\vspace{1em}


\section{Introduction}
\label{sec:introduction}

% Context
In modern application development following an agile process, application source code is constantly being changed, including the source code of Function-as-a-Service (FaaS) applications.
Due to the pay-per-use model of FaaS where usage is billed on millisecond accuracy, whenever an update introduces a performance regression the execution also becomes more expensive.
In very large scale FaaS applications comprising multiple functions with billions of requests per month, every small performance change can have a drastic impact on the total cost of running the application.

% Problem
Detecting if a function code adjustment introduces a relevant performance change usually requires developers to deploy both source code versions of the respective function separately, stress them with an artificial load, and use the measurements to decide if the updated new function version introduces a performance change.
Using this traditional benchmarking approach in FaaS environments is hard, because there are various random factors and parameters which can not be specified by a developer~\cite{paper_leitner_cloud_variability, paper_bermbach2017_expect_the_unexpected, schirmer_night_2023}. 
It thus requires a large amount of requests to average out all random effects and get accurate results.

\begin{figure}
    \centering
    \includegraphics[width=0.7\columnwidth]{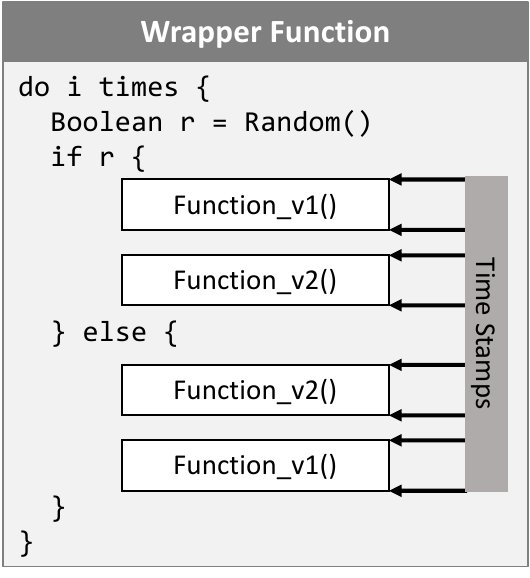}
    \caption{The wrapper function applies the RMIT execution order. It runs both function versions successively multiple times and measures each execution.}
    \label{img:approach}
\end{figure}

% Idea
In this paper, we propose to apply the Randomized Multiple Interleaved Trials (RMIT) execution methodology~\cite{abedi_conducting_2017} in FaaS benchmarking and use bootstrapping percentile intervals to more accurately detect performance changes with fewer requests.
We enable RMIT using an automatically generated wrapper function which includes both function versions, executes both versions successively multiple times, and detects performance changes based on injected measurement statements (see \cref{img:approach}). 
This ensures that both function versions are executed in the same environment, on the same executor instance, and competing for the same CPUs at this time.
They thus share the same environment including other functions running on this instance and all the other random factors influencing the experiment are close to equal.
The injected measurement statements track both starting and end timestamps of all function executions and enable an accurate performance evaluation without numerous benchmarking requests.
This completely automated process can be leveraged in continuous integration and deployment (CI/CD) pipelines to automatically detect performance regressions.

% Contributions
\newpage
Our contributions in this regard include:
\begin{enumerate}
    \item A function-wrapper approach which applies the RMIT execution strategy to detect performance changes in FaaS applications more efficiently.
    \item An open source prototype which enables an automatic detection of performance changes in FaaS function source code using the RMIT execution methodology. 
    \item An extensive study comparing RMIT and the traditional benchmarking methodology on AWS Lambda (AWS) and Google Cloud Functions (GCF) using function source code which includes an artificial and adjustable performance issue.
\end{enumerate}

% Results
Our results show that the RMIT execution methodology can drastically reduce the confidence interval width of the results and thus enables a more fine-grained detection of performance changes. 
In our study, a benchmark configuration using $150$ measurement pairs in total is sufficient to detect the injected performance change accurately (max $1.7\%$ confidence interval width using RMIT, compared to up to $7\%$ using traditional benchmarking with $150$ measurements for each function version).
This finding motivates to prefer and apply the RMIT benchmarking methodology in CI/CD pipelines of FaaS applications to detect performance changes as soon and accurately as possible.

\section{Background}
\label{sec:background}

This paper describes an approach to automatically measure performance metrics of FaaS applications.
We use benchmarking to determine the execution latencies of functions and then apply a statistical bootstrapping method to detect performance changes between successive function versions.

% FaaS
\free
\textbf{Function-as-a-Service (FaaS)}
is a service model in which developers divide their software into individual functions which are then usually deployed on a cloud provider and triggered by certain events, e.g., a user request~\cite{paper_bermbach2021_cloud_engineering}. 
The cloud provider handles all involved aspects such as the deployment, dynamic scaling, or event handling; software developers have only a limited influence on the deployment, they can, e.g., define the deployment region or the size of memory~\cite{paper_bermbach2020_faas_coldstarts}.

% Benchmarking in cloud environments
\free
\textbf{Benchmarking} 
evaluates non-functional requirements such as latency by stressing the system under test, a deployed function in our case, with an artificial load.
Benchmarking in cloud environments is a challenging task, as there are many random factors and fluctuations that affect the experiment run.
To counteract this, an experiment must be repeated several times to collect meaningful results~\cite{book_bermbach2017_cloud_service_benchmarking}.

% Bootstrapping
\free 
\textbf{Bootstrapping} 
percentile intervals uses a hierarchical random re-sampling with replacement of measured values to derive confidence intervals for two measurement series.
We use this bootstrapping methodology to determine the confidence intervals of measured performance changes in our study as it does not assume an underlying distribution of measurement values~\cite{kalibera2020quantifying, hesterberg2015what}.
\section{Efficient Function Benchmarking}
\label{sec:approach}

\begin{figure}
    \centering
    \includegraphics[width=0.8\columnwidth]{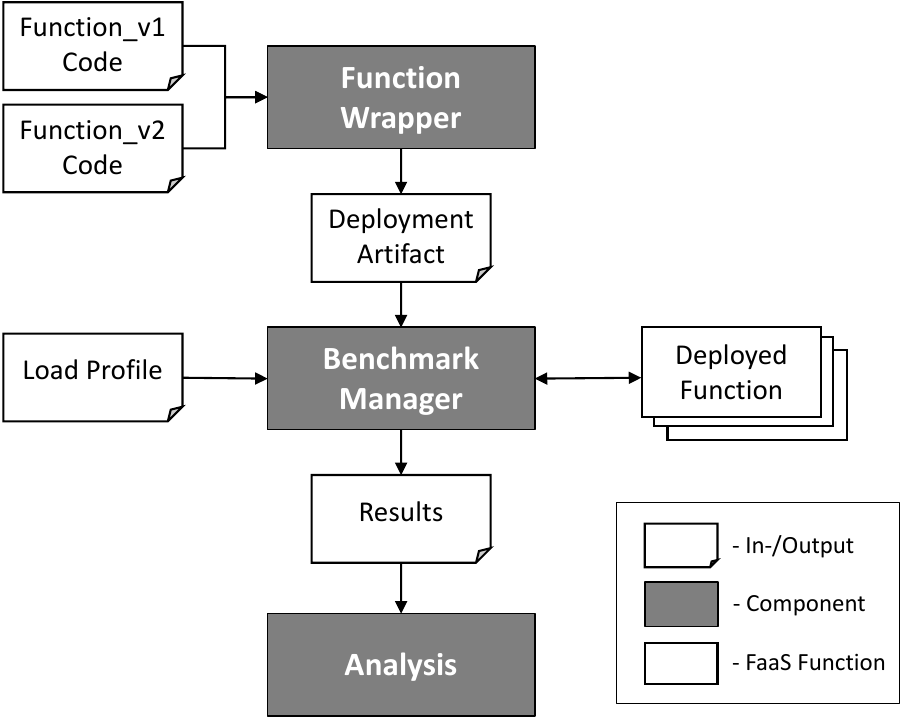}
    \caption{The function wrapper component automatically merges both function versions into a single function. The benchmark manager then deploys this function several times, runs a workload against the function endpoints, and forwards the results to the analysis component, which finally calculates the performance change using bootstrapping.}
    \label{img:process}
\end{figure}

\Cref{img:process} illustrates the general function benchmarking process.
First, the function wrapper component creates the deployment artifact which contains both function versions, the respective time stamping statements, and the iterative loop which calls both functions. 
Second, the benchmark manager deploys this artifact multiple times and executes a workload which equally stresses all wrapper functions.
Finally, the analysis uses a bootstrapping algorithm to determine the performance change with confidence intervals.

\free
\textbf{1) Wrapping Functions:}
The function wrapper component merges the source code of both function versions into a single function which implements the RMIT execution methodology and calls both versions successively multiple times. 
To decide which function is executed first (version 1 followed by version 2 or the other way around), the wrapper function uses a random value (see \cref{img:approach}).
The additional loop which executes the functions multiple times ensures more stable results.
During the benchmark experiment, it is important to mix both execution sequences to mitigate warm up effects and the random fluctuations in the cloud environment.
Furthermore, the function wrapper also automatically places time stamp statements before and after each function call to enable a fine-grained measurement of each internal function execution time. 
Finally, the function wrapper produces a single wrapper function as a deployment artifact.

Two challenging tasks during this merge are (i) to substitute and deal with calls to the data layer and (ii) to ensure that both function versions use their respective imported library version (we discuss this in \cref{sec:discussion} in detail).

\free
\textbf{2) Run Benchmark:}
The benchmark manager first deploys the artifact several times and stores the function endpoints. 
This ensures that the wrapper function is benchmarked on multiple executor instances on the FaaS provider and thus is increasing the validity of our results, because more influencing factors are taken into account.
Next, the manager runs a load profile against all endpoints equally to stress the functions.
Finally, the benchmark manager collects all function execution measurements and creates a result data set containing all function version 1 and version 2 execution times.

\free
\textbf{3) Analysis:}
The analysis component calculates the median performance change and the respective confidence intervals for this change based on the result set.
In this study, we use the median measurement of all measurement pairs to determine the median performance change and use bootstrapping percentile intervals to calculate the 99\% confidence intervals. 
\section{Experimental Evaluation}
\label{sec:evaluation}
We evaluate our wrapper approach in a study comparing it to the traditional benchmarking in which both functions are benchmarked individually. 
For this, we use our open source prototype to study a function that contains a configurable performance change. 
In the study, we inject different performance change levels, deploy the functions on two public cloud FaaS providers, and evaluate the detection capabilities of both benchmarking methodologies.

\free
\textbf{Open Source Prototype:}
Our open source implementation\footnote{\url{https://github.com/martingrambow/faasterBench}} implements the function wrapper, benchmark manager, and analysis component. 
The final wrapper function can be configured to use one of three execution modes.
The first executes follows the RMIT execution mode, the second and third mode only executes version 1 or version 2 to enable a comparison between our wrapper approach and the traditional one. 
Furthermore, it is possible to adjust the load profile, e.g., the number of iterations or function invocations.

\free
\textbf{Studied Function:}
We use a function which estimates Pi using the Monte Carlo method.
The algorithm works by generating multiple points on a $1$x$1$ square.
It then calculates the distance to the center of coordinates, and sorts the points on whether they are located in the circle or not. 
The ratio between both groups is the estimation of Pi.
As the estimation with more generated values becomes better and better, but also increases the computing time linearly, the number of values/points is a good parameter to study the performance change. 

\free
\textbf{Study Design:}
We deploy the Pi estimation function in the AWS eu-central-1 region and in the europe-west3 GCF region. 
In each setup, we deploy the wrapper function $5$ or $10$ times to ensure multiple executor instances.
In each experiment, we generate $5,000,000$ values for version 1 of the function.
For version 2, we either leave it the same, or increase it by $5\%$ (to $5,250,000$ values).
The $0\%$ regression represents an A/A test to evaluate if the approach successfully does not detect a performance change, the $5\%$ regression represents the introduction of a performance regression. 
Furthermore, we use load profiles of $5$, $10$, and $25$ calls to each deployed function.
In the RMIT execution mode, we use $i=3$ for the number of iterations.
To enable a fair comparison with the traditional approach, we repeat each call to the traditionally executed functions three times.
In total, we thus have $75$ measurement pairs for the smallest setup ($5$ wrapper functions x $5$ calls per function x $3$ iterations) and $750$ ($10$ x $25$ x $3$) for the largest one.

\begin{figure}[t]
    \centering
    \includegraphics[width=\columnwidth]{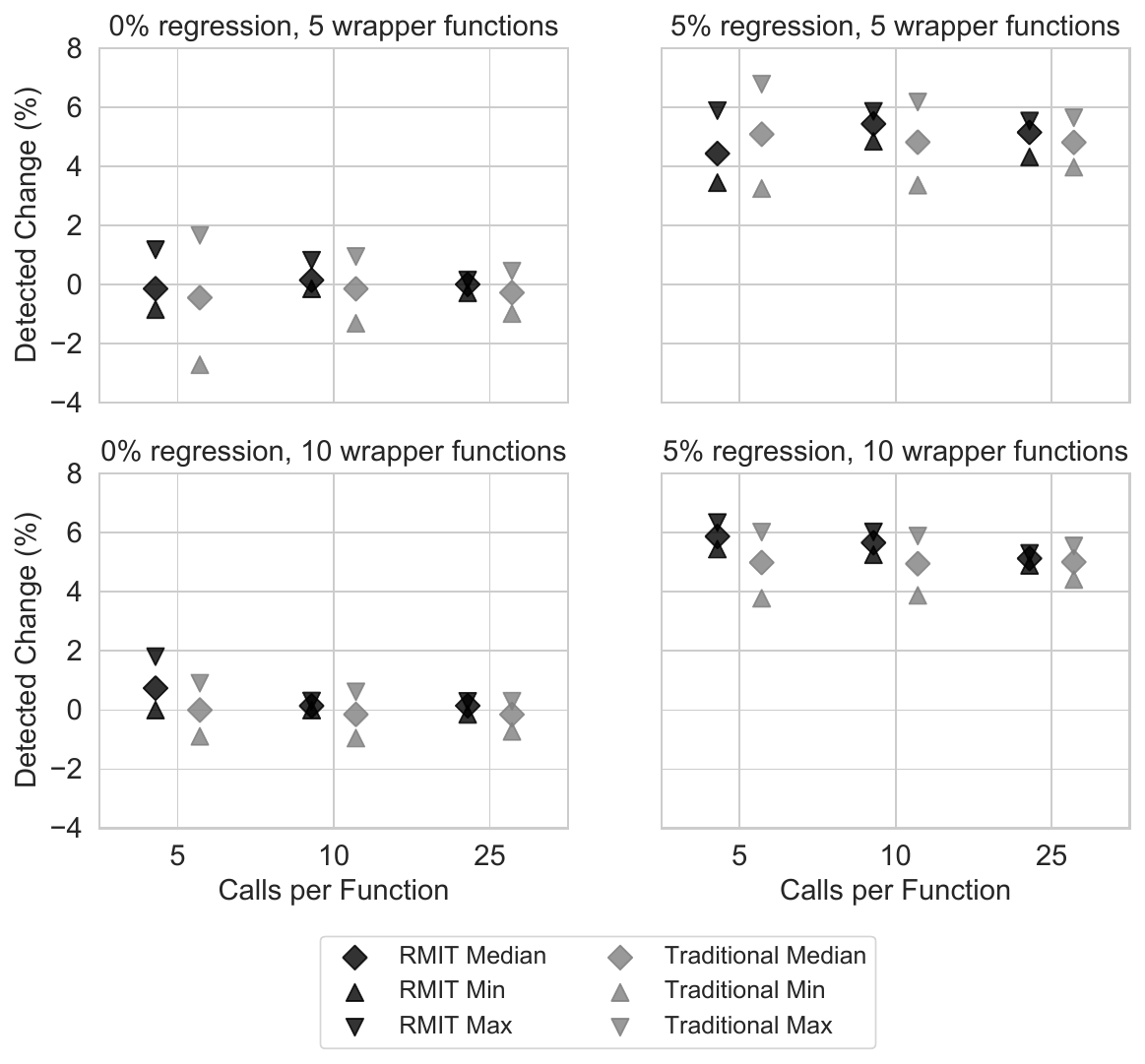}
    \caption{For the AWS FaaS provider, RMIT and the traditional approach correctly detect no performance regression in the A/A test experiment and raise alarms for the $5\%$ injected regression. RMIT, however, report smaller CI width and is thus more efficient.}
    \label{img:resultsAWS}
\end{figure}

\begin{figure}[t]
    \centering
    \includegraphics[width=\columnwidth]{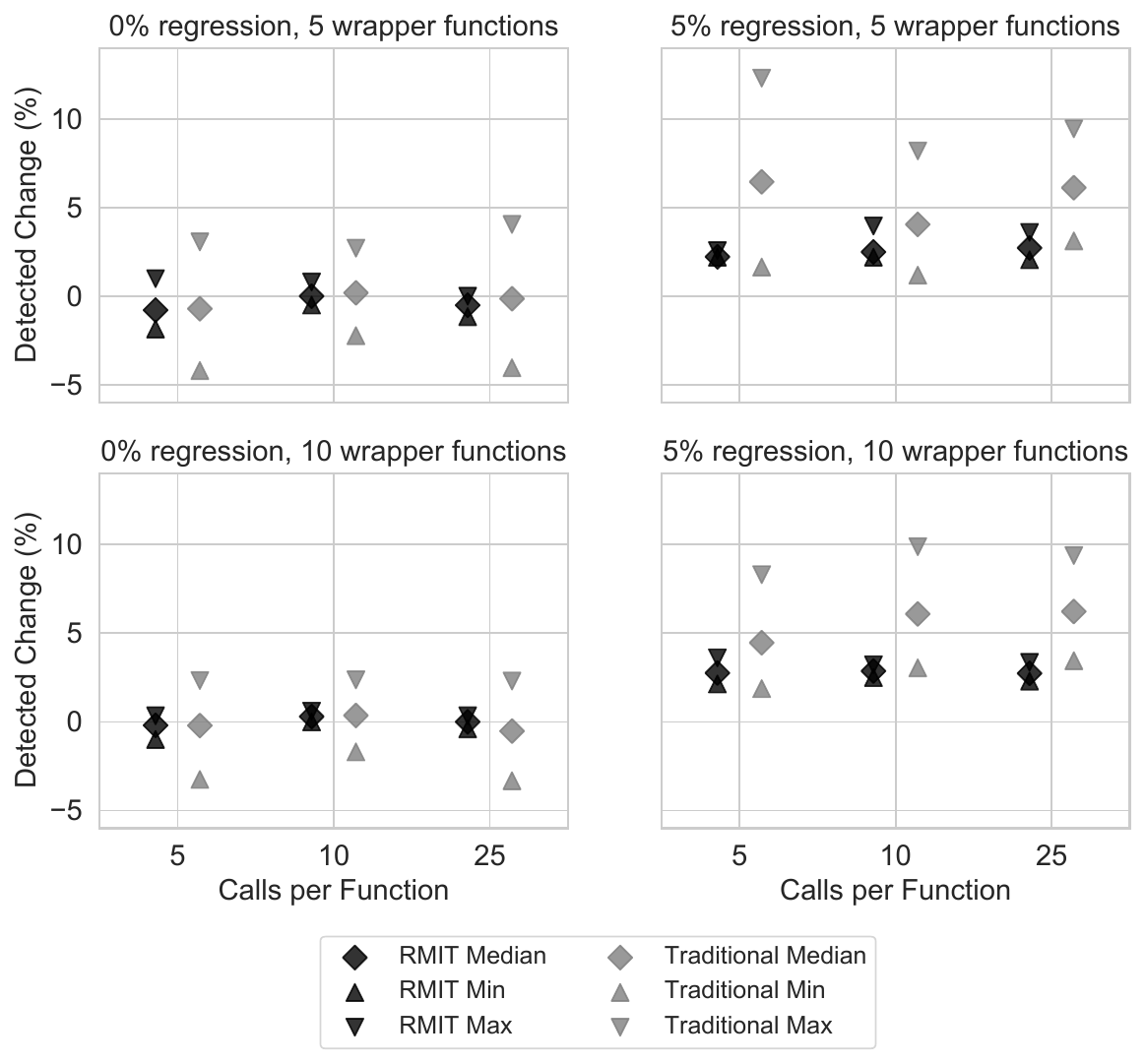}
    \caption{The RMIT methodology is more efficient for the GCF provider as well. The detected performance change, however, does not match the injected $5\%$ regression value.}
    \label{img:resultsGoogle}
\end{figure}

\free
\textbf{Results:}
\Cref{img:resultsAWS,img:resultsGoogle} illustrate the results of our study.
In both studied FaaS platforms and all evaluated benchmark configurations, all confidence intervals (CI) straddle the $0\%$ performance mark in the A/A test and indicate a performance change when configuring the studied function to calculate $5\%$ more values.
Thus, both approaches, RMIT and the traditional approach, correctly detect no performance change in the A/A test, but raise an alarm with the injected regression.
The RMIT approach, however, is more efficient and reports way smaller CIs. 
It is able to detect performance changes more accurately and with fewer requests.

For AWS (see \cref{img:resultsAWS}), using RMIT shrinks the CI width by more than $2.75$ times (from $2.81\%$ to $1.02\%$ using $5\%$ regression, $5$ wrapper functions, $10$ calls per function).
While $5$ calls per function can still result in CI widths of up to $2.4\%$ for the RMIT approach (traditional $4.4\%$), both other evaluated parameters result in more stable results. 
Our experiments report a maximal CI width for the RMIT approach of $1\%$ using $10$ calls per function (traditional: $2.8\%$) and $1.2\%$ using $25$ calls (traditional: $1.6\%$).
Thus, $10$ calls per functions seem to be a good trade-off between benchmarking effort and accuracy for AWS as more function calls do not result in better accuracy.

The traditional benchmarking approach reports even larger CI widths on GCF (see \cref{img:resultsGoogle}).
Here, the largest RMIT improvement shrinks the CI width from $10.65\%$ to $0.37\%$, which translates to only about $3\%$ of the CI width using the traditional approach ($5\%$ regression, $5$ wrapper functions, $5$ calls per function).
Furthermore, the maximal CI width of $2.83\%$ using the RMIT wrapper is still smaller than the smallest CI width of $4\%$ using the traditional approach.
In total, using $10$ calls per function is for GCF a good trade-off between accuracy and benchmarking effort as well, as it results in a CI width of max $1.7\%$ for our experiments and more function calls barely improve the accuracy.

Finally, the detected performance change value varies between both FaaS providers. 
While the $5\%$ injected regression are more or less accurately detected on AWS, while GCF only detects a $3\%$ change using the RMIT methodology.

\free
\textbf{Key Findings:}
RMIT enables a more fine-grained detection of performance changes compared to the traditional benchmarking approach as it reports smaller CI widths. 
In our study using an artificial performance regression in realistic cloud benchmarking environments, a benchmark configuration of $3$ RMIT iterations, $5$ deployed wrapper functions and $10$ calls per wrapper function is sufficient to detect the injected performance change accurately ($150$ measurement pairs in total) for both studied FaaS providers.
The detected performance change value, however, does not have to match the actual regression value.

\section{Discussion}
\label{sec:discussion}

The RMIT benchmarking methodology can be used in FaaS applications to detect introduced performance changes in individual functions more efficiently.
Our study showed that RMIT enables an accurate detection with only $150$ measurement pairs in total, even in a cloud environment with multiple random factors impacting the benchmark experiment results.
While these results are promising, applying the RMIT wrapper approach to actual production FaaS application still requires several improvements and features which we want to discuss further.

\free
\textbf{Dependency Updates:}
Common source code updates include upgrading used external libraries to the newest version. 
Currently, our prototype implementation does not support different versions of libraries to enable a comparison between two function versions which only differ in the imported library versions. 
While different library versions can also be compared using standard benchmarking techniques outside of FaaS platforms, 
we plan to enable and study these dependency updates in future work.

\free
\textbf{Data Modifications:}
Our current study object is a compute-only function without any calls to external services or data layer. 
Applying the RMIT wrapper approach in real FaaS applications, however, requires a feature which deals with these requests in a fair way.
For example, a list of customer orders is already sorted after the first function call -- the succeeding RMIT function calls will finish sooner and falsify the results.
Thus, we plan to introduce a feature multiplying the function parameter as well and, e.g., forward six customer IDs to enable a sorting of six customer order lists.
Moreover, we plan to include a detailed tracing of external calls and exclude these durations from the function execution time.

\section{Related Work}
\label{sec:related}
This paper motivates to include a dedicated continuous benchmarking step in CI/CD pipelines of software artifacts to detect performance issues as soon as possible.
While our study uses bootstrapping and the median value to determine the performance change, there are further algorithms which can be applied in the result analysis as well.
Finally, there are multiple related papers dealing with benchmarking in FaaS environments.

\free
\textbf{Continuous Benchmarking:}
Several research papers already motivate a dedicated performance evaluation step in integration and deployment pipelines of software to detect performance regressions~\cite{foo_mining_2010, foo_industrial_2015, paper_grambow2019_continuous_benchmarking, daly_industry_2019, ingo_automated_2020, daly_creating_2021}.
For FaaS applications, however, there are only a few studies dealing with function updates~\cite{lee_evaluation_2018}.
Comparable to the automated benchmark process in our study, there are several related approaches and tools to automate benchmark experiments in cloud environments~\cite{silva_cloudbench_2013, paper_hasenburg2021_mockfog2, grambow_befaas_2021}.
To the best of our knowledge, we are the first who apply and study a dedicated automated continuous benchmarking step in FaaS environments to detect performance changes. 

\free
\textbf{Performance Change Detection:}
Alternative and complementary algorithms for determining the performance change value and detecting regressions use threshold values, evaluate performance signatures, apply machine learning, and automatically adjust to cloud variability~\cite{paper_grambow2019_continuous_benchmarking, foo_mining_2010, daly_industry_2019, toslali_iter8_2021, grambow2022microbenchmark}.
Studying these approaches in combination with the RMIT methodology can potentially improve the detection capabilities further.

\free
\textbf{Benchmarking FaaS:}
Related research dealing with FaaS studies specific features such as cold start latencies, instance lifetimes, tail latencies, or performance fluctuations~\cite{wang_peeking_2018, manner_cold_2018, shahrad_architectural_2019, schirmer_night_2023, ustiugov_analyzing_2021}.
The findings of these studies can help to improve the RMIT benchmark configuration parameters and analysis of results.
\section{Conclusion}
\label{sec:conclusion}

This paper describes a function wrapper approach which enables the RMIT benchmark execution methodology in FaaS applications and compares its detection capabilities with the traditional benchmarking approach.
Applying RMIT can drastically reduce the confidence interval width of results and thus enables a more fine-grained detection of performance changes in function code, which directly influences cost. 
Our study shows that RMIT can be used in automatic CI/CD pipelines of FaaS applications to detect performance changes more efficiently -- $150$ measurements are sufficient to reliably and accurately detect an artificially injected performance regression in a cloud environment.

\balance
\bibliographystyle{unsrt}
\bibliography{bibliography}

\end{document}